\documentclass[aps,prd,preprint,groupedaddress]{revtex4}

\usepackage{graphicx}

\begin{document}

\title{New analyses of double-bang events in the atmosphere}

\author{C\'elio A. Moura}
\email[]{moura@ifi.unicamp.br}
\altaffiliation[Also at ]{Departamento de F\'{\i}sica, Centro de
  Investigaci\'on y de Estudios Avanzados del IPN. Apdo. Postal 14-740
  07000 M\'exico DF, Mexico.} 
\author{Marcelo M. Guzzo}
\email[]{guzzo@ifi.unicamp.br}
\affiliation{Instituto de F\'{\i}sica Gleb Wataghin - UNICAMP\\
	Cx.P.6165 13083-970 Campinas, SP Brazil}

\date{\today}

\begin{abstract}
We use CORSIKA+Herwig simulation code to produce ultra-high energy
neutrino interactions in the atmosphere. Our aim is to reproduce
extensive air showers originated by extragalactic tau-neutrinos. For
charged current tau-neutrino interactions in the atmosphere, beside
the air shower originated from the neutrino interaction, it is expected
that a tau is created and may decay before reaching the ground. That
phenomenon makes possible the generation of two related extensive air showers, the so called Double-Bang event. We make an analysis of
the main characteristics of Double-Bang events in the atmosphere for
mean values of the parameters involved in such phenomenon, like the
inelasticity and tau decay length. We discuss what may happen for the
``out of the average'' cases and 
conclude that it may be possible to observe this kind of event in
ultra-high energy cosmic ray observatories such as Pierre Auger or
Telescope Array.
\end{abstract}


\maketitle

\section{Introduction}

It is believed that neutrinos with energies of the order of
$10^{18}$~eV (1~EeV) arrive at the Earth from extragalactic sources
and may interact with the nuclei in the atmosphere generating cascades
of particles called Extensive Air Showers (EASs)~\cite{sokolsky,eas}. There
are several techniques to measure EASs, but for energies above 1~EeV basically two of them are being used: one is to detect the
particles when they reach the ground through water tanks equipped with
photomultipliers~\cite{arrays,Yamamoto:2006iq,Letessier-Selvon:2005uq}. The photomultipliers are used to
detect the Cherenkov light emitted by the ultra relativistic particles
when they cross the water inside the tanks. By spreading the Cherenkov
tanks in a big and plane area it is possible to determine the
distribution of particles when the EAS reach the ground and with this
information to determine the direction and energy of the incident
particle that generated the event. The other technique is to observe the
scintillation light of the atoms which are excited during the
developing process of the
EAS~\cite{Yamamoto:2006iq,Letessier-Selvon:2005uq,fluor}. The atoms
emit detectable visible and ultraviolet light when they return to their
ground states. The detectors used in this method are called
Fluorescence Detectors (FDs) because of the similarity of the
scintillation with the fluorescence light. This technique of detection
can be used to measure the profile of the EAS when it develops through
the atmosphere.

To distinguish which particle have generated an EAS is not an easy
task, but we can take advantage of the different depth of the
atmosphere depending on the zenith angle. While for vertical angles
the atmospheric depth is approximately 1000~g/cm$^2$, for horizontal
angles it is approximately 36000~g/cm$^2$. Ordinary
cosmic rays like protons, electrons, photons, helium nucleus or heavier nuclei, interact in the top of the atmosphere and when they come from almost horizontal angles the EAS generated arrives at the detector basically only with the muonic component, because the hadronic component is extinguished and the electromagnetic one is almost completely absorbed by the atmosphere~\cite{zas,bertou}.
Neutrinos, on the other hand, because of their
very low cross section, are the only particles beside the
muons that can go deep in the atmosphere before interacting. In this way it is possible that neutrinos coming from very
near horizontal angles interact close to the detector.

The case of tau-neutrinos is even more special. Because
it is not expected that tau-neutrinos are created in extragalactic
sources, i.e., tau-neutrinos are due to neutrino oscillation during the
propagation from the source to the
detector~\cite{learned,athar}, and, what is the subject of this paper, because when the tau-neutrino interacts via charged current (CC) in the atmosphere
it creates an EAS which contains a tau. For energies of the order of
1~EeV, the tau decays in a distance comparable to the size of the EAS
and almost all of the times it generates another EAS that may also be
detectable. This kind of unique signature, with two EASs
coming from the same direction in a time interval of the order of the
tau mean lifetime, is named Double-Bang event.

The detection of a Double-Bang would be very important both from the astrophysical and the particle physics point of view. Because this event, as already mentioned, has a unique Ultra-High Energy (UHE) tau-neutrino signature given by two EASs coming from the same direction in a time interval corresponding to the tau decay length, with an approximately determined energy relation, it can be used to test the existence of the UHE tau-neutrinos. On the other hand the non observation of Double-Bang events together with the Earth Skimming~\cite{bertou,Miele:2005bt,Aramo:2004pr,Bottai:2002nn,gora} events can help to set an upper limit for the UHE tau-neutrino flux. The future data of the Pierre Auger Observatory~\cite{opa} for inclined EASs, that means EASs coming from zenith angles bigger than $60^\circ$, will be of fundamental importance on this analysis. The suppression of the cosmic ray flux above the GZK cutoff does not apply for neutrinos however the source anisotropy recently measured~\cite{Cronin:2007zz} 
can indicate some nearby UHE neutrino sources, and theoretical models that relate the fluxes of UHE neutrinos and cosmic rays can be tested, as well as UHE neutrino oscillations.

We present simulations
of Double-Bang events generated in the atmosphere by 
tau-neutrinos with energies of 0.5~EeV, 1.5~EeV and 5~EeV. The
inelasticity for the simulated events is approximately 0.2, 0.73, and
0.92 for each energy respectively, the tau decay length is of 20~km
and the energy fraction from the tau that goes into the second EAS is of 2/3.
We also discuss cases with different energies, decay lengths,
inelasticities, and second EAS's energy. Because for ultra-high
energies the neutrino cross section, inelasticity, and flux are still
uncertain, and because in our analysis we do not use any specific
detector characteristic, we do not calculate the Double-Bang event
rate for neutrinos interacting in the atmosphere. The characteristics
we observe from the simulations confirm that Double-Bang events can
develop in the atmosphere and FDs of experiments like Pierre
Auger~\cite{Verzi:2007eg}, HiRes and Telescope Array~\cite{Jui:2006da}
may detect them depending on the neutrino flux and cross section, as
it was pointed out in~\cite{Guzzo:2005fi}.

\section{Ultra-High Energy Neutrino Induced Events}

Consider a neutrino arriving at the Earth and interacting via CC with
a nucleon in the atmosphere generating a charged lepton and other
fragments:

\begin{equation}
\nu_l+N\to l+X \,,
\end{equation}
where $l$ is the lepton flavor ($e$, $\mu$, $\tau$). For a neutral
current (NC) interaction another neutrino is created rather then a
charged lepton.

After the CC interaction, the neutrino energy is divided between the
charged lepton and the other fragments that generate the EAS in the
following way:

\begin{equation}
E_\nu=E_1+E_l \,,
\label{eq:energy}
\end{equation}
where $E_\nu$ is the incident neutrino energy, $E_1$ is the energy
deposited in the fragments that generate the EAS and $E_l$ is the
charged lepton energy.

Because the energy distribution varies for each interaction, it is
interesting to define the inelasticity, that is the fraction of the
neutrino energy that goes to the EAS and not to the charged
lepton. The inelasticity is:

\begin{equation}
y=(E_\nu-E_l)/E_\nu \,.
\label{eq:inelast}
\end{equation}

Combining Eqs.~(\ref{eq:energy}) and~(\ref{eq:inelast}) we have:

\begin{equation}
E_1=yE_\nu \,,
\label{eq:eshower}
\end{equation}
and finally from Eqs.~(\ref{eq:energy}) and~(\ref{eq:eshower}) we find
the neutrino energy fraction transfered to the charged lepton:

\begin{equation}
E_l=(1-y)E_\nu \,.
\label{eq:e2ndbang}
\end{equation}

When a muon-neutrino interacts via CC, it creates an EAS with the same
characteristics of the ones produced via NC interactions for any
neutrino flavor. It is because similarly to the neutrinos created
after the NC interactions, the muon created after a CC muon-neutrino
interaction almost does not interact with the atmosphere~\cite{pdg1}
and for the energies considered, the muon decay length is much longer
than its interaction length.

Electron and tau-neutrino CC interactions are completely
different. The electron, created after the electron-neutrino
interaction, interacts immediately generating a cascade of
electromagnetic particles beside the hadronic component generated by
the other fragments created in the first interaction. The tau created
after the tau-neutrino interaction propagates in a very similar way to
the muon, but its mean lifetime is much shorter. For the energies
considered the decay length of the tau is of the order of few tens of
km in the laboratory frame, comparable to the length of the hadronic
EAS created. When the tau decays, it may generate a second hadronic
shower and both showers together generate a sign characteristic of
tau-neutrino interactions only.

\section{Tau-Neutrinos Interacting in the Atmosphere\label{sec:tau}}

Many articles have been published analyzing the potential of the
Pierre Auger Observatory to detect and investigate 
neutrinos~\cite{e-mu-nu,exotic} and specially Earth
Skimming events. In reference~\cite{gora} down-going
tau-neutrino events are also investigated. According to the estimated
neutrino flux coming from Active Galactic Nuclei~\cite{agnsn,agntds},
Gamma Ray Bursts~\cite{grb}, Topological Defects~\cite{agntds,tds},
and Cosmogenic Neutrinos ~\cite{Allard:2006mv,Engel:2001hd} 
as well as to the neutrino cross section extrapolated to ultra-high
energies~\cite{nudis}, it is expected a number of tau-neutrino events
in Auger approximately of the order of between 0.1 and 1 event per
year, with energies around 1~EeV. 

Based on the propose of John Learned and Sandip Pakvasa
to detect tau-neutrinos with energies of the order of 1~PeV in
detectors under water or ice~\cite{learned}, it has been proposed to
use FDs to detect tau-neutrinos with energies of the order of 1~EeV
interacting in the atmosphere~\cite{Guzzo:2005fi,Guzzo:2003pu}.
Here we investigate through simulations the properties of the
Double-Bang events in the atmosphere. These properties do not depend
on the detector, despite we are concerned with the characteristics as
efficiency and field of view for FDs. 

After it is created the tau propagates on average, before decaying in the
laboratory frame, a distance given by:

\begin{equation}
\langle L\rangle=\gamma c\tau \,,
\end{equation}
where $\gamma$ is the Lorentz factor and $\tau$ is the tau mean
lifetime. In terms of the energy given in units of EeV, we have that
the distance traveled by the tau in the laboratory frame is, on average:

\begin{eqnarray}\label{lmed1}
\langle L\rangle&\simeq&\frac{E_\tau}{\mbox{[EeV]}}\times49~\mbox{km} \,, \\
&\simeq&(1-y)\frac{E_\nu}{\mbox{[EeV]}}\times49~\mbox{km} \,.
\label{lmed2}
\end{eqnarray}

In our simulations we consider only the case of hadronic decay,
despite we could have considered also the electronic decay ($\tau\to
e\nu_e\nu_\tau$), which corresponds to almost 18\% of the tau decay
branching 
ratio~\cite{pdg2} and produce a electromagnetic shower with, on
average, about 1/3 of the tau energy. Only the hadronic branching
ratio is responsible 
for 64\% of the tau decays. Based on the hadronic branching ratio of
the tau decay, that is basically decay to pions, and considering that
each decay product carries on average the same amount of energy, we
calculate that the 
energy of the second EAS, originated by the tau decay, is on average:

\begin{eqnarray}\label{e2tau}
\langle E_2\rangle&\approx&\frac{2}{3}E_\tau \,, \\
&\approx&\frac{2}{3}(1-y)E_\nu \,.
\label{e2nu}
\end{eqnarray}

This expression is confirmed by simulations~\cite{tauola}.

\section{Simulations}

Due to the low event rate expected for the neutrinos, combined with the limitations of the detection and discrimination techniques, there is no measured UHE
EAS that could be uniquely associated to a primary neutrino. Monte Carlo simulations are an important
tool to study the characteristics of UHE events that may be generated
by neutrinos. In general, the EAS longitudinal development depends on the
energy and the type of the primary particle, as well as on the
interaction depth and the incident angle in the atmosphere. Until this moment, the only way to study in a systematical
way the longitudinal development of the
charged particles in a Double-Bang event is through simulations.

A study of the characteristics of the UHE EASs generated by electron
and muon-neutrinos was made in~\cite{Ambrosio:2003nr} through
CORSIKA~\cite{Heck:1998vt} simulations. CORSIKA versions 6.203 and
6.204 themselves, that we used to do the simulations here, are not
able to 
simulate neutrinos as primary particles, so the neutrino interaction
is made by Herwig~\cite{Corcella:2000bw} and then, the results of this
interaction are taken by CORSIKA as the primary particles which give
rise to the EAS.
In CORSIKA, the hadronic interaction models used were QGSJET~\cite{Ostapchenko:2005nj} 
for high energies and GHEISHA~\cite{Fesefeldt:1985yw} 
for low energies ($<10^{11}$eV). We used a thinning of $\varepsilon_{th}=10^{-6}$, what permits the program to follow all the particles with energies down to 10~GeV. The energy cutoffs were 0.05, 0.05, 0.0001, and 0.0001~GeV for hadrons, muons, electrons and photons respectively.
Until the present moment, simulations with 
tau-neutrino as primary particle and the decay of taus in
CORSIKA are not available. Because of that we simulate
Double-Bang events through phenomenological arguments, using muon
neutrinos as primary particles and pions as the generators of the EAS
which in principle may come from the tau decay. A more accurate and particular
analysis for Auger, e.g., can be done using the method described in~\cite{gora},
specially for the simulation of tau decays with the software
TAUOLA~\cite{tauola} and the light propagation and the hardware
detector trigger simulated by means of the Auger software framework
called Offline~\cite{offline}. We want to be more general and to do
simulations specific for Auger is beyond the scope of this paper. 

\begin{figure}[htbp]
   \includegraphics*[width=\textwidth]{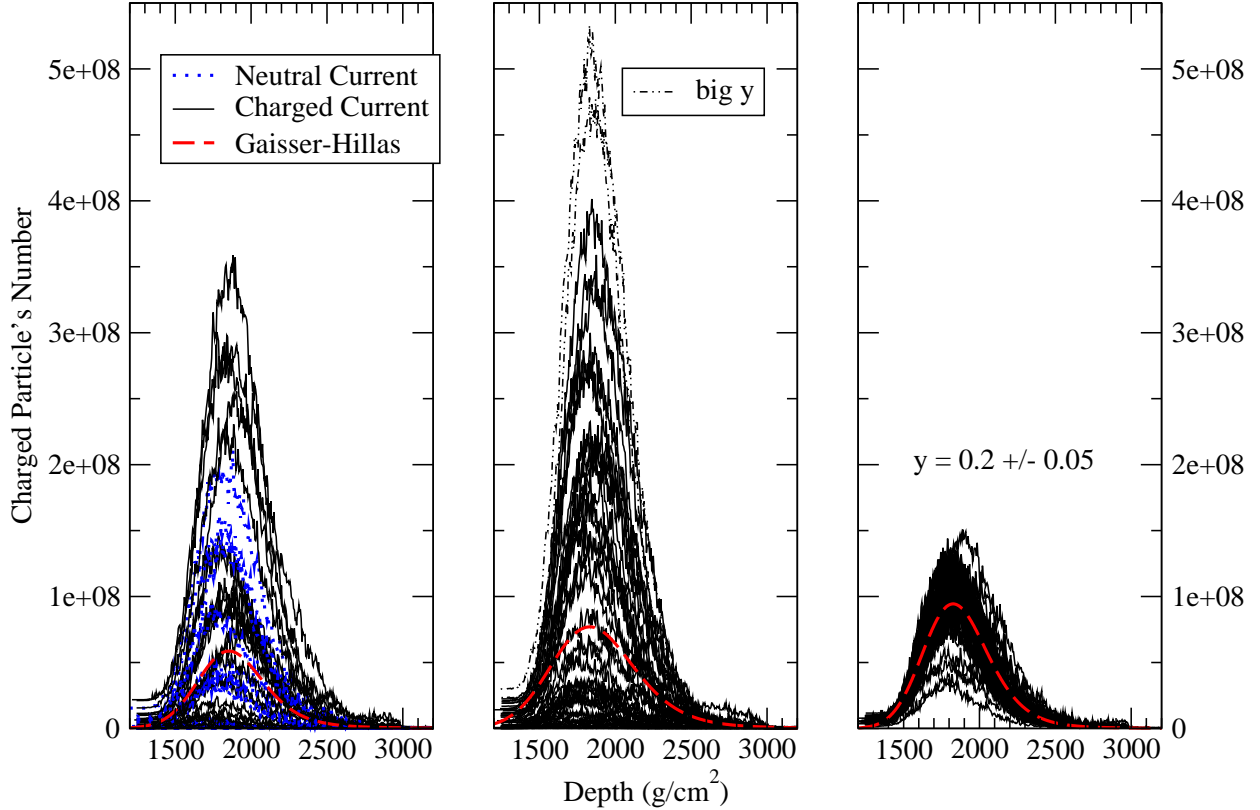} 
   \caption{Charged particles as a function of atmospheric
     depth. Comparison of the kind of interaction of muon-neutrino
     induced events with energy $E=0.5$~EeV. In the left side we have
     interactions randomly induced via Monte Carlo. In the center only
     CC interactions are present, and in the right side we have CC
     interactions with inelasticity $y=0.2\pm0.05$. Read text for
     details. \label{fig:compar}} 
\end{figure}

Fig.~\ref{fig:compar} presents the different behavior of an EAS depending
on the type of interaction and the inelasticity value. In each of the
graphics of the figure we have 50 EASs induced by muon-neutrino with
energy $E=0.5$~EeV and incident angle of $75^\circ$. The dashed 
lines are the Gaisser-Hillas
fit~\cite{Gaisser:1977sa} to the average of all the EASs in a graphic. The
discussion here can be thought to be naive, but it is very useful to
show that the simulation code is producing, on average, what is
expected in the framework of the Standard Model. 

On the left side of the Fig.~\ref{fig:compar} one has CC and NC
interactions happening freely. Also the elasticity value is obtained
randomly by Monte Carlo method with CORSIKA. We obtained 35 CC
interactions (black continuous lines) and 15 NC interactions (blue
dotted lines). As it is expected because of the bigger cross section,
70\% of the interactions are CC interactions.

On the central part we present only CC interactions. There are two
showers with bigger number of charged particles than in the
simulations presented on the left side, but it may probably be because
of statistical fluctuations. On average, for tau and muon-neutrino
induced EASs, one expects the same longitudinal profile for both CC
and NC interactions. The difference as we have seen, beside the cross
section difference between CC and NC, is that in the case of CC a
charged lepton is created and for NC it is a neutrino.

The right side of the Fig.~\ref{fig:compar} has only EAS's generated
via CC interaction and with inelasticity $y=0.2\pm0.05$. Because the
most part of the energy goes to the charged lepton, the EAS with
hadronic nature has low energy. Moreover, because the inelasticity is
basically fixed around one value, the maximum number of charged
particles is concentrated around a fixed value too. For that value of
inelasticity, the number of charged particles is around $10^8$
particles. The Gaisser-Hillas function gives a good approximation with
low deviation in this case. We can see that, as expected for these
energies~\cite{cteq6}, for the two previous cases the average value of
$y$ is nearly the same, but with much bigger deviation from the
average since the inelasticity is a free parameter chosen randomly by
the simulation code. 

\subsection{Double-Bangs}

The alternative we use to simulate events generated by tau-neutrinos
in the atmosphere is to input muon-neutrino as primary particle. The
main characteristics of muon and tau-neutrino induced events are the
same with the exception that a muon-neutrino interacting via CC
generates a muon. As we are simulating tau-neutrino interactions, we
need the creation of a tau. To simulate the tau creation and
subsequent decay we use pions. One can divide the tau decay in two
main modes: 64\% for the hadronic modes and 36\% for the leptonic
modes. Despite the leptonic decay mode may generate a electromagnetic
shower 18\% of the times as we discuss in Section~\ref{sec:tau}, in
our simulations we consider only the hadronic modes in which 
the main decay products are pions. In this way we generate the second
shower inputing the pion as primary particle in the same direction of
the muon-neutrino, after a distance $L$, from
Eqs.~(\ref{lmed1})-(\ref{lmed2}), corresponding to the tau mean 
lifetime and with the corresponding energy $E_2$ from
Eqs.~(\ref{e2tau})-(\ref{e2nu}).

\begin{figure}[htbp]
   \includegraphics[angle=-90,width=\textwidth]{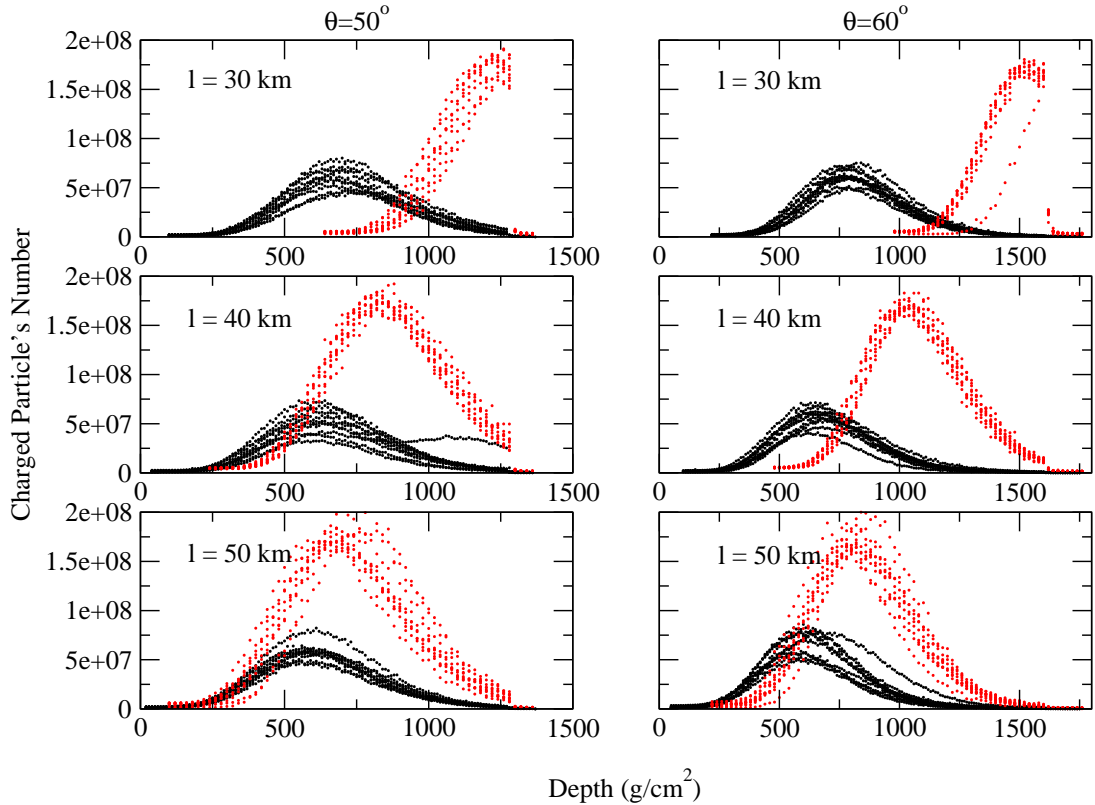} 
   \caption{Double-Bangs generated by tau-neutrinos in the
     atmosphere. Each graphic contains ten events represented by the
     number of charged particles as a function of depth with primary
     neutrino energy of 0.5~EeV.  The incident angle is of 50$^\circ$
     in the first column and 60$^\circ$ in the second one. The
     interaction distance to the ground is $l=30$, 40 and 50~km
     respectively for each 
     line. The first EASs, with less than $10^8$ charged particles,
     are represented in black and the second ones, due to the tau
     decay with more than the double number of particles, are
     represented in red.\label{fig:db1}}
\end{figure}

\begin{figure}[htbp]
   \includegraphics[angle=-90,width=\textwidth]{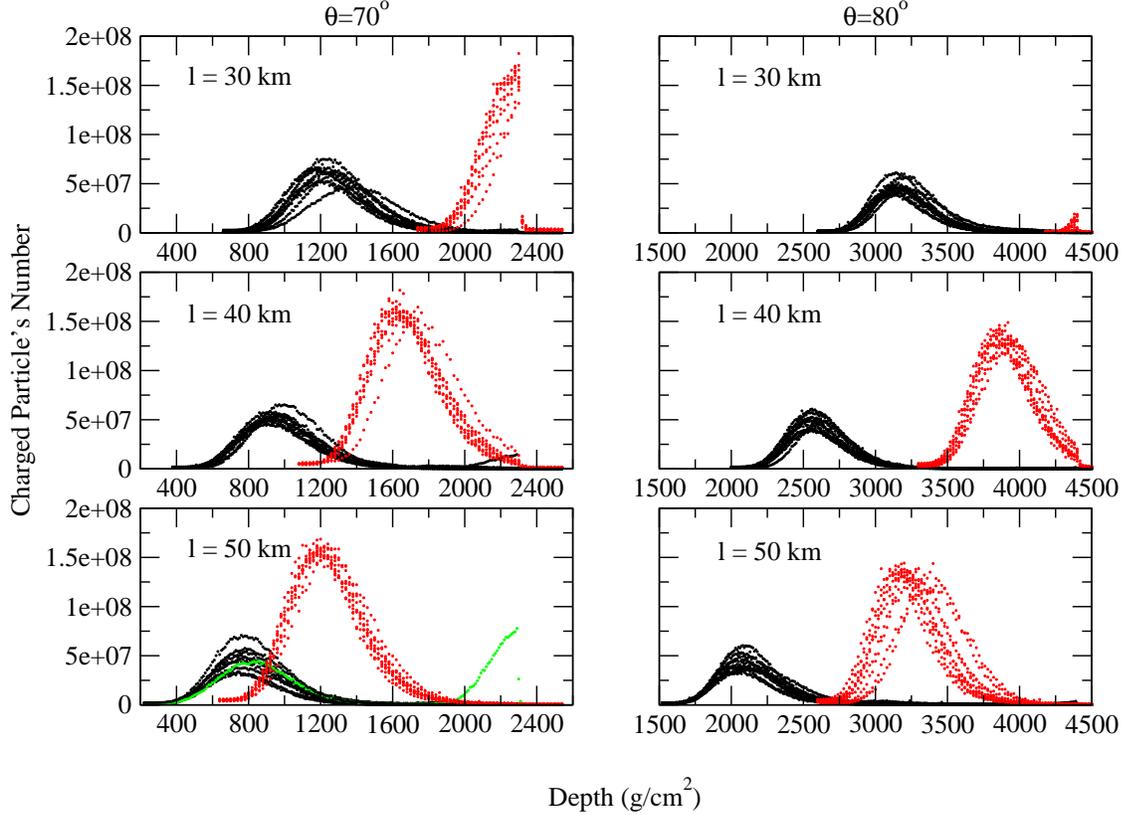} 
   \caption{Double-Bangs generated by tau-neutrinos in the
     atmosphere. Each graphic contains ten events represented by the
     number of charged particles as a function of depth with neutrino
     energy of 0.5~EeV. The incident angle is of 70$^\circ$ in the
     first column and 80$^\circ$ in the second one. The interaction
     distance to the ground is $l=30$, 40 and 50~km respectively for
     each line. The first 
     EASs, with less than $10^8$ charged particles, are represented
     in black and the second ones, due to the tau decay with more than
     the double number of particles, are represented in
     red. There is a fake Double-Bang event represented in green in
     the bottom left.\label{fig:db2}} 
\end{figure}

\begin{figure}[htbp]
   \includegraphics[angle=-90,width=\textwidth]{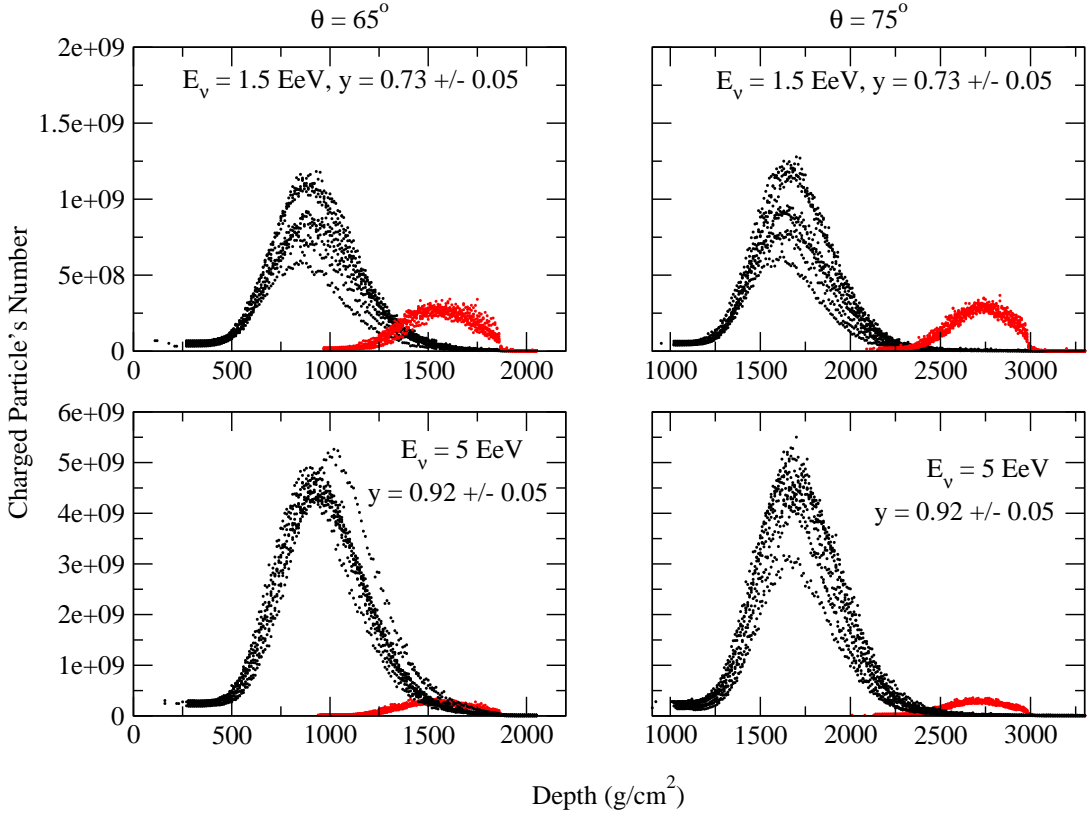}
   \caption{Double-Bangs generated by tau-neutrinos in the
     atmosphere. Each graphic contains ten events represented by the
     number of charged particles as a function of depth generated
     35~km far from the detector. The incident angle is of 65$^\circ$ in
     the first column and 75$^\circ$ in the second one. The neutrino
     energy is 1.5~EeV for the graphics in the first line and 5~EeV
     for the ones in the second line. The first EASs, with bigger
     number of charged particles, are represented in black while the
     second ones, due to the tau decay, are represented in
     red.\label{fig:db3}}
\end{figure}

In Figs.~\ref{fig:db1},~\ref{fig:db2}, and~\ref{fig:db3} we show
events for the tau energy of approximately 0.4~EeV. It means that on average the
tau should have run a distance of about 20~km from the neutrino
interaction point until the tau decay, and considering, from
Eq.~(\ref{e2tau}), that 2/3 of the tau energy goes to the decay
products, it corresponds to a pion energy of 0.27~EeV. Each graphic of
the figures contain 10 Double-Bang events, that is to say 10 showers
initiated by muon-neutrino and 10 initiated by pion ($\pi^-$). The
distance, in the neutrino propagation axis, from the neutrino
interaction point to the ground is represented by $l$. 

In the bottom left of Fig.~\ref{fig:db2} we have a kind of fake
Double-Bang event where the second EAS was not generated by the tau
decay, but by some fragment of the first EAS that interacted or
decayed after traveling a very long distance comparing with the mean
distance traveled by the taus with 0.4~EeV energy. We checked that this second generated EAS is of electromagnetic characteristic and it was probably generated by a muon decay. It is important to notice that among the 310 simulations initiated by neutrino that we made, in only one it happened to be produced a second shower imitating
the second bang of a Double-Bang event.  It would be a background of less than 0.5\% of the total number of events. Adding to that, the fake Double-Bang is out of
the average and with an electromagnetic second Bang, what completely distinguish them from the case we are studying with hadronic second Bangs.

To obtain the events generated by 0.5~EeV neutrinos depicted in
Figs.~\ref{fig:db1} and~\ref{fig:db2} we use only inelasticity values
of $y=0.2\pm0.05$. In this way we have, from Eq.~(\ref{eq:eshower}),
$E_1\sim E_\nu/5$, and also from Eq.~(\ref{e2nu}),
$E_2\approx2.67E_1$. Analyzing events coming from different angles and
interaction depths, we may say which values of theses parameters are
more favorable for detecting Double-Bang events with energies of about
0.5~EeV and 20\% of inelasticity. 

In Fig.~\ref{fig:db3} we have events with more energy for the primary
neutrino. In this case, due to the energy distribution between the
first EAS and the charged lepton, we have the first EAS more energetic
than the second one. For $E_\nu=1.5$~EeV and $y=0.73$, $E_1=1.1$~EeV
$\approx4E_2$. If the neutrino energy is $E_\nu=5$~EeV and
$y\approx0.92$, than $E_1=4.6$~EeV $\approx17E_2$.

Observing the simulations in Figs.~\ref{fig:db1}
and~\ref{fig:db2} we have the hint that the bigger is the incident
angle, for neutrinos interacting always at the same distance from the
detector, the bigger is the distance between the two maximums of the
generated EASs. Furthermore, the same happens if the
distance from the primary interaction to the detector diminishes. We
conclude that 
the worse situations to detect Double-Bangs generated by neutrinos
with energies of the order of 0.5~EeV and inelasticity of about 20\%
are for $l=50$~km and 
$\theta=50^\circ$, as well as for $l=30$~km and $\theta=80^\circ$. In
the first situation the two EASs overlap each other and in the second
one, the second EAS reach the ground before it has a considerable
number of particles. These effects may happen due to the different
atmospheric density depending on the height. Close to the surface of
the Earth, where the density is higher, the particles have more
probability of interacting and consequently the longitudinal
development of the EAS is faster.

\section{Discussion and Summary}

UHE neutrinos probably coming from extragalactic sources may interact
in the atmosphere via NC or CC. The mean inelasticity for energies
of the order of 1~EeV for neutrino-nucleon interaction is
$\langle y\rangle\approx0.2$. It means that for NC interactions one
hadronic EAS is 
produced on average with approximately 20\% of the energy of the
primary neutrino while for CC interactions there are three distinct
cases for each neutrino flavor. 
We analyze the case when a tau-neutrino interacts and a tau is created
which decay in a distance comparable to the size of the hadronic EAS
generated by the neutrino interaction. Nearly 64\% of the times the
tau decay generates a hadronic EAS and 18\% of the times an
electromagnetic one. Approximately 17\% of the times the tau decay in
muon and neutrinos that do not generate any cascade.

We simulate tau-neutrino induced events for three different energies
with $y$ values such that Double-Bang events have optimal
characteristics to be observed. In the 
case of $E_\nu=0.5$~EeV the mean inelasticity is compatible with
the observation of the events because the average tau energy is
$\langle E_\tau\rangle\simeq0.4$~EeV what corresponds to a mean
distance travelled before 
decaying of $\langle L\rangle\simeq20$~km. For this specific values of
$E_\nu$, $y$, and $L$ the EAS that may be 
generated by the tau decay develops very close to the EAS created by
the neutrino-nucleon interaction, but at the same time both EASs are
distinguishable. For the simulated energies of $E_\nu=1.5$~EeV and
$E_\nu=5$~EeV, if we consider just the mean inelasticity value, i.e.,
$y=0.2$, the energy of 
the tau is such that the tau decay does not happen inside the field of
view of a supposed detector because on average it travels too long
distance compared to the field of view 
before decaying. So we simulate events with values of $y$ above the
average for the tau energy to correspond on average to a travel distance of
$L\simeq20$~km.

\begin{figure}[htbp]
   \includegraphics[angle=0,width=\textwidth]{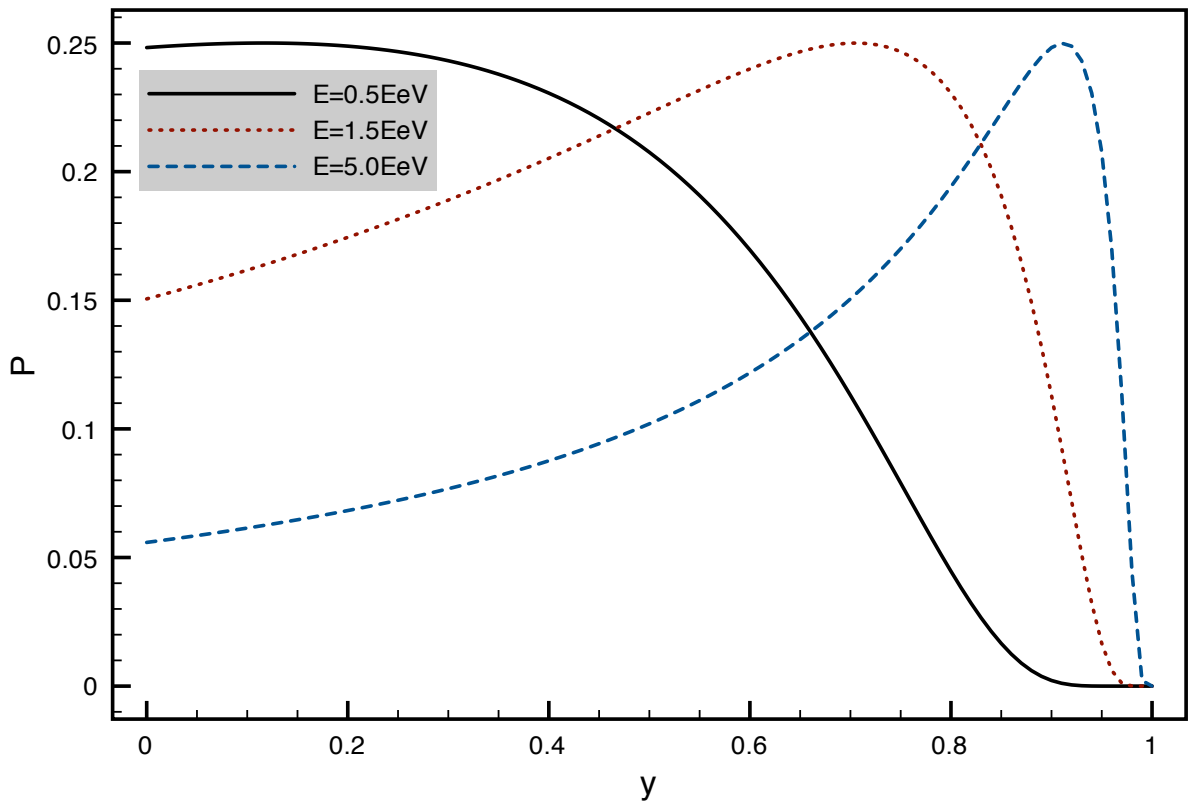}
   \caption{Probability for the tau decay in a distance between 15 and
     30~km after being created as a function of the inelasticity, for
     different values of primary neutrino energy.\label{fig:compar2}} 
\end{figure}

Therefore, we have made for this paper simulations for specific cases
of primary neutrino energy and inelasticity, and used the mean values
of tau decay length and the energy from the tau that goes to the
second hadronic shower in a Double-Bang event. What happens if we
consider the general case when the average values can be the not real
or not measured ones? In Fig.~\ref{fig:compar2} we plot the
probability for the tau decay in a distance between 15 and 30~km after
being created as a function of the inelasticity, for the different
values of primary neutrino energy used in the simulations. Despite the
mean inelasticity for these energies is approximately 0.2, it has a
very big dispersion. As can be seen in Fig.~7 of ref.~\cite{gora},
about 10\% of the events are expected to be in the range of
inelasticity $y=0.2\pm0.05$, 45\% for $y<0.1$, and 30\% for
$y>0.3$. One sees from Fig.~\ref{fig:compar2} that the simulations we
present are for the cases of inelasticity when the probability of the
tau decay in the range between 15 and 30~km is larger. Because
approximately 70\% of the events occur with inelasticity lower than
0.3, among the cases analyzed in this paper, the most probable events
are generated by neutrinos with energy of $E_\nu=0.5$~EeV. 

It is important to notice that the $y$ distribution of the cross
section is very dependent on the low-$x$ behavior of the parton
distributions in a region beyond the access of accelerator
experiments~\cite{CastroPena:2000fx}, so if the $x$ behavior is not
the expected by the extrapolation of the parameters of the Standard
Model, the $y$ dispersion may be different and our conclusions too. The Double-Bang events have the energy of the two EASs directly related to the inelasticity, as can be seen in Eqs.~(\ref{eq:eshower})-(\ref{eq:e2ndbang}), and the detection os these kind of events can be of great importance to study the behavior of $y$ at ultra-high energies.

If the Double-Bang event characteristics like inelasticity and tau
decay length follow the average, the energy range of the primary
neutrino must 
be very strict because to observe the two EASs the primary neutrino
energy cannot 
exceed about 1~EeV. But also the efficiency of the detector is an
important factor and energies bellow 1~EeV are not optimal for the
FDs to observe events. Because the dispersion specially in the
inelasticity and tau decay length, the primary neutrino energy range
can be relaxed. The simulations we present give us a sign that the
profile of Double-Bang events may be observed by FDs such
as those of the Pierre Auger Observatory or Telescope Array. We
also suggest that ground arrays such as the one of the Telescope Array 
and the Auger array may be another possibility to detect
Double-Bangs. Almost horizontal Double-Bangs may develop both EASs
inside the arrays. And finally the other possibility could be to use both
techniques, for example, observing the first shower with the FD and the
second, due to the tau decay, with the ground array.

We have studied the parameter space in which Double-Bang events may be measured by FDs. Taking the expected average values for the inelasticity and tau decay length at ultra-high energies, if the neutrino interaction happens between 30 and 50~km far from the detector, the incident angle must be bigger than $50^\circ$ and the most favorable to detect the events is around $75^\circ$ (see Figs.~\ref{fig:db1}, \ref{fig:db2}, and~\ref{fig:db3}). From Fig.~\ref{fig:compar2} we learn that considering the inelasticity for the neutrino DIS cross section around 0.2, the primary neutrino energy around 0.5~EeV is optimal to detect Double-Bangs. In this sense, the HEAT extension of the Auger experiment~\cite{Klages:2007zza} and the TALE extension of the Telescope Array experiment~\cite{taleprop} will be of great importance also in the study of Double-Bangs. Both experiments will have bigger efficiency and field of view for events with energies bellow 1~EeV.

\begin{acknowledgments}
We thank specially O. Pisanti for her help with the simulations and
discussions. The work was supported by CNPq, FAPESP and AlBan.
\end{acknowledgments}

\bibliography{papsim.bib}

\end{document}